\documentclass[10pt,final,journal,twocolumn]{IEEEtran}
\usepackage[utf8]{inputenc}
\usepackage{cite}
\usepackage[cmex10]{amsmath}
\usepackage{amsmath}
\usepackage{amsfonts}
\usepackage{amsthm}
\usepackage{algorithm}
\usepackage{multirow}
\usepackage{multicol}
\usepackage{stfloats}
\usepackage{multicol,multienum}
\usepackage{multirow}
\usepackage[caption=false,font=footnotesize]{subfig}
\usepackage{wrapfig}
\usepackage{color}
\usepackage{graphicx}
\usepackage{xspace}
\usepackage{epstopdf}
\usepackage{suffix}
\usepackage{mathtools}
\usepackage{bm}

\DeclarePairedDelimiterX\MeijerM[3]{\lparen}{\rparen}%
{\begin{smallmatrix}#1 \\ #2\end{smallmatrix}\delimsize\vert\,#3}

\newcommand\MeijerG[8][]{%
  G^{\,#2,#3}_{#4,#5}\MeijerM[#1]{#6}{#7}{#8}}

\WithSuffix\newcommand\MeijerG*[7]{%
  G^{\,#1,#2}_{#3,#4}\MeijerM*{#5}{#6}{#7}}

\begin{document}

\title{Uncertainty of Resilience in Complex Networks with Nonlinear Dynamics}

\author{Giannis Moutsinas,
        Mengbang Zou$^*$,
        Weisi Guo,~\IEEEmembership{Senior Member,~IEEE}
        
\thanks{G. Moutsinas is with Cranfield University, Cranfield MK43 0AL, U.K.  (e-mail:
giannismoutsinas@gmail.com).
}
\thanks{M. Zou is with Cranfield University, Cranfield MK43 0AL, U.K. (e-mail: M.Zou@cranfield.ac.uk)}
\thanks{W. Guo is with Cranfield University, Cranfield MK43 0AL, U.K., and also with the Alan Turing
Institute, London, NW1 2DB, U.K. (e-mail: weisi.guo@cranfield.ac.uk).}}

\markboth{}%
{Shell \MakeLowercase{\textit{et al.}}: }

\maketitle{}

\begin{abstract}
Resilience is a system's ability to maintain its function when perturbations and errors occur. Whilst we understand low-dimensional networked systems' behaviour well, our understanding of systems consisting of a large number of components is limited. Recent research in predicting the network level resilience pattern has advanced our understanding of the coupling relationship between global network topology and local nonlinear component dynamics. However, when there is uncertainty in the model parameters, our understanding of how this translates to uncertainty in resilience is unclear for a large-scale networked system. Here we develop a polynomial chaos expansion method to estimate the resilience for a wide range of uncertainty distributions. By applying this method to case studies, we not only reveal the general resilience distribution with respect to the topology and dynamics sub-models, but also identify critical aspects to inform better monitoring to reduce uncertainty.
\end{abstract}

\begin{IEEEkeywords}
Uncertainty; Resilience; Dynamic Complex Network
\end{IEEEkeywords}

%
\IEEEpeerreviewmaketitle

\section{Introduction}
%
%
%
%
\IEEEPARstart{O}{rganized} behavior in economics, infrastructure, ecology and human society often involve large-scale networked systems. These systems network together relatively simple local component dynamics to achieve sophisticated system wide behaviour. A critical part of the organized behavior is the ability for a system to be resilient the ability to retain original functionality after a perturbation of failure. A system’s resilience is a key property and plays a crucial role in reducing risks and mitigating damages \cite{ref1, ref2}. Research on resilience of dynamic network has arisen in lots of areas and has widespread applications including service disruption in communication systems caused by terminal failures \cite{ref3}, blackout in power systems caused by power station shutdowns \cite{ref4}, the loss of biodiversity caused by the decline in ecology \cite{ref5}. Whilst we understand low-dimensional models with a few interacting components well \cite{ref2}, our understanding of multi-dimensional systems consisting of a large number of components that interact through a complex network is limited . Recent research in predicting the network-level\cite{ref6} and node-level resilience pattern\cite{ref7} has advanced our understanding of the coupling relationship between topology and dynamics. 


\quad To simulate the dynamics and estimate resilience of complex networks with dynamical effects, we need to define dynamical models with parameter values. However, in practice, uncertainty on the model form and parameters are inherently present. Uncertainty can originate from latent process variables (process noise), e.g., inherent biological variability between cells which are genetically identical \cite{ref8} or from a parameter estimation procedure based on noisy measurements (measurement or inference noise). For example, a recent research proposed an analytical framework for exactly predicting the critical transition in a complex networked system subjected to noise effects \cite{ref9}. In recent years, the modeling and numerical simulation of practical problems with uncertainty have received unprecedented attention, which is called Uncertainty Quantification (UQ). UQ methods have been applied in widespread fields like fluid dynamics \cite{ref10}, weather forecasting \cite{ref11}, etc. At present, UQ methods are shown as follows \cite{ref12}: 

\subsection{Review of Uncertainty Quantification}

\quad Monte Carlo Methods \cite{ref13} are based on samples. In these methods, samples are randomly generated according to probability distribution. For each sample, the problem to be solved becomes a definite problem. By solving these determined problems, representative statistical information about the exact solution can be discovered. These methods are easy to use, but need large sample data. For arbitrarily large dynamical networks, it is difficult to sample appropriately without a foundation UQ theory. 

\quad Perturbation Methods \cite{ref14} expands a function into a Taylor series around its mean value, and then make a reasonable truncation. Normally, at most we can truncate the second-order expansion, because for higher-order cases, the resulting solution system will become very complicated.

\quad Moment Equation Methods \cite{ref15} attempt to directly solve the equations satisfied by the moments of the random solution. These equations about moments need to be derived from the original stochastic problem. For some simple problems, such as linear problems, this method is more effective. But usually, when we derive a certain moment equation, we need to use the information of higher moments. 

\quad Polynomial approximation method \cite{ref16} is a standard method for UQ in singular dynamical systems. The basic idea is to perform polynomial expansion of the exact solution in a random parameter space. This method could solve problems with any type of random parameter inputs. First, we need to perform a finite order expansion of the exact solution in the random parameter space and then take this expansion into the original problem and do Galerkin projection in the expansion polynomial space. After that we get a simultaneous equations about the expansion coefficient. By solving the equations, we can get all the statistical information of the exact solution. If the exact solution has good regularity for random parameters and this method can achieve exponential convergence.


\subsection{Contribution}

\quad The contribution of this paper is to take uncertainty into account when estimating resilience of dynamic networks. Even though recent research about resilience of network is prevalent, research in this area considering uncertainty is lacking. In practical problems, not taking this uncertainty into account possibly leads to deviation when estimating resilience of a system. Therefore, considering uncertainty when estimating resilience of dynamic complex network have great significance.

\quad In this paper, we propose a method with polynomial chaos expansion to quantify these uncertain factors to reduce the risk of uncertainty when estimating the resilience of dynamic network. And then, we analysis how parameters and network topology with uncertainty affect the resilience of dynamic network, which would give us more insight of dynamic network.


\section{System Setup}

\subsection{Saddle-node bifurcation}
\quad The traditional mathematical treatment of resilience used from ecology \cite{ref17} to engineering \cite{ref18} approximates the behavior of a complex system with a one-dimensional nonlinear dynamic equation
\begin{equation}\label{equ1}
\dot{x}=f(\beta,x)
\end{equation} 

\begin{figure*}[t]
\centering
\resizebox*{16cm}{!}{\includegraphics{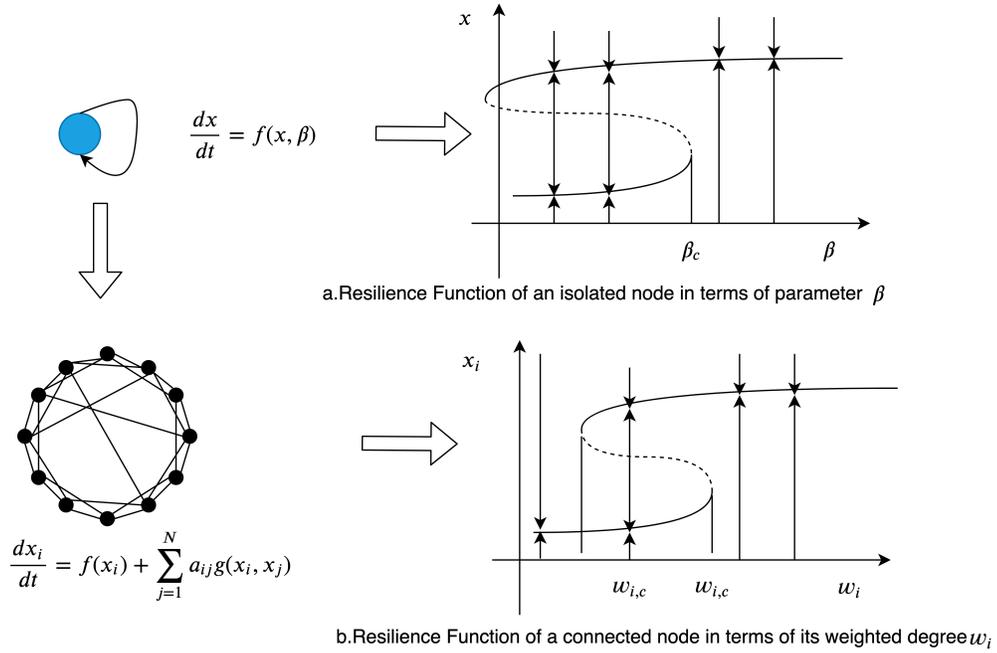}}
\caption{It shows dynamics of a single node and the coupled dynamics in a complex network}
\label{fig:1}
\end{figure*}

\quad The functional form of \(f(\beta,x)\) represents the system's dynamics, and the parameter \(\beta\) captures the changing environment conditions (show in Figure \ref{fig:1}). The system is assumed to be in one of the stable fixed points, \(x_0\) of equation (\ref{equ1}), extract from
\begin{equation}\label{equ2}
    f(\beta, x_0)=0
    \end{equation}
    \begin{equation}\label{equ3}
        \left. \frac{df}{dx} \right|_{x=x_0}<0
    \end{equation}
where equation (\ref{equ2}) provides the system's steady state and equation (\ref{equ3}) guarantees its linear stability.

\quad The saddle-node or fold bifurcation is which two equilibria of a dynamical system collide and annihilate each other. The simplest example of such bifurcation is\begin{equation}\label{equ4}
    \dot{x} = x^2 - c 
\end{equation}
If \(c>0\), then there are 2 equilibria, stable one at $-\sqrt{x}$ and unstable one at $\sqrt{x}$. If $c<0$, there are no equilibria for the system since $x^2-c$ is always positive. For $c=0$, we have the bifurcation point and only one equilibrium exists, which is not hyperbolic.

\begin{figure}
\centering
\subfloat[A non-resilient system.]{%
\resizebox*{8cm}{!}{\includegraphics{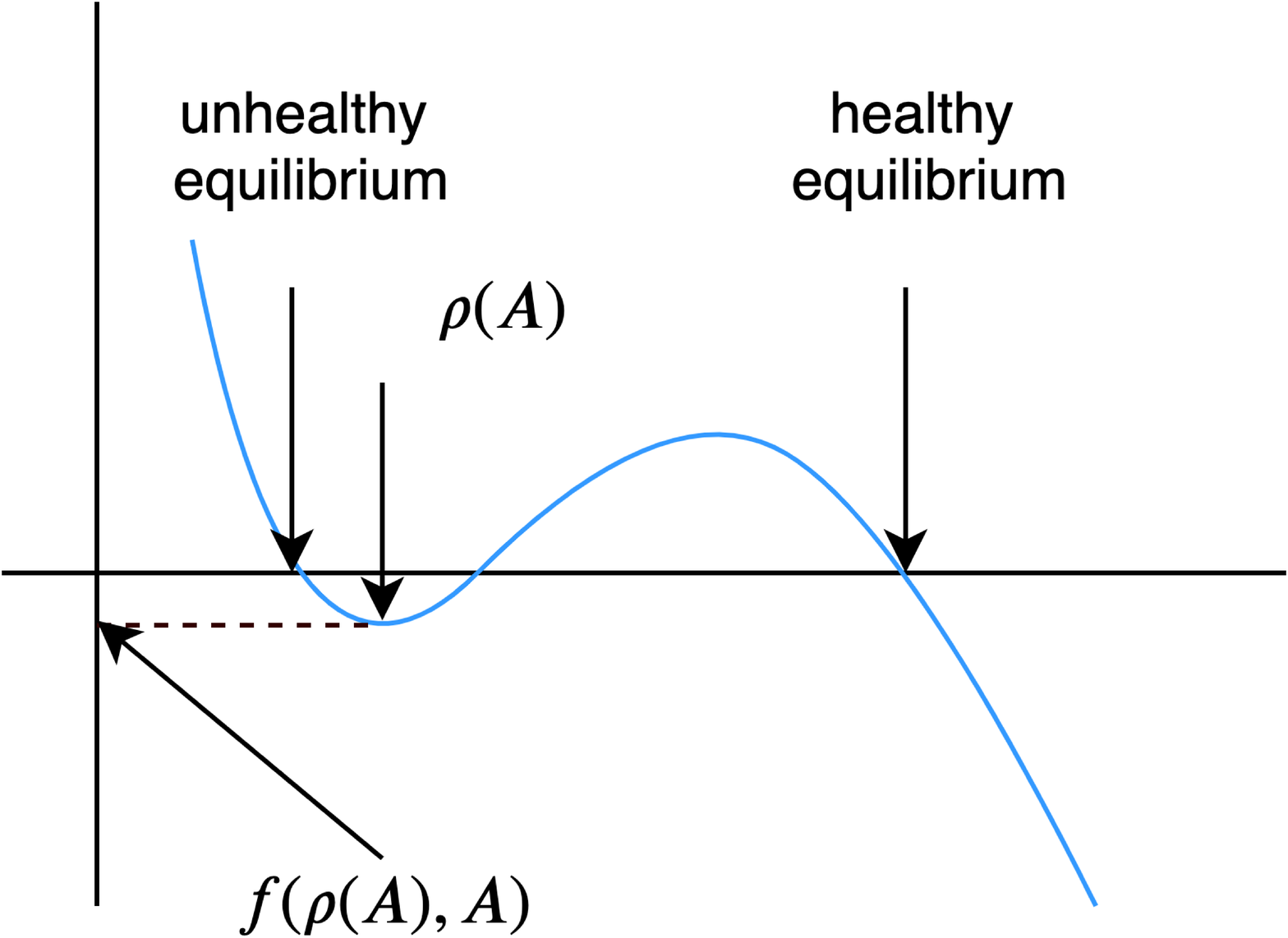}}}\hspace{5pt}
\subfloat[A resilient system]{%
\resizebox*{8cm}{!}{\includegraphics{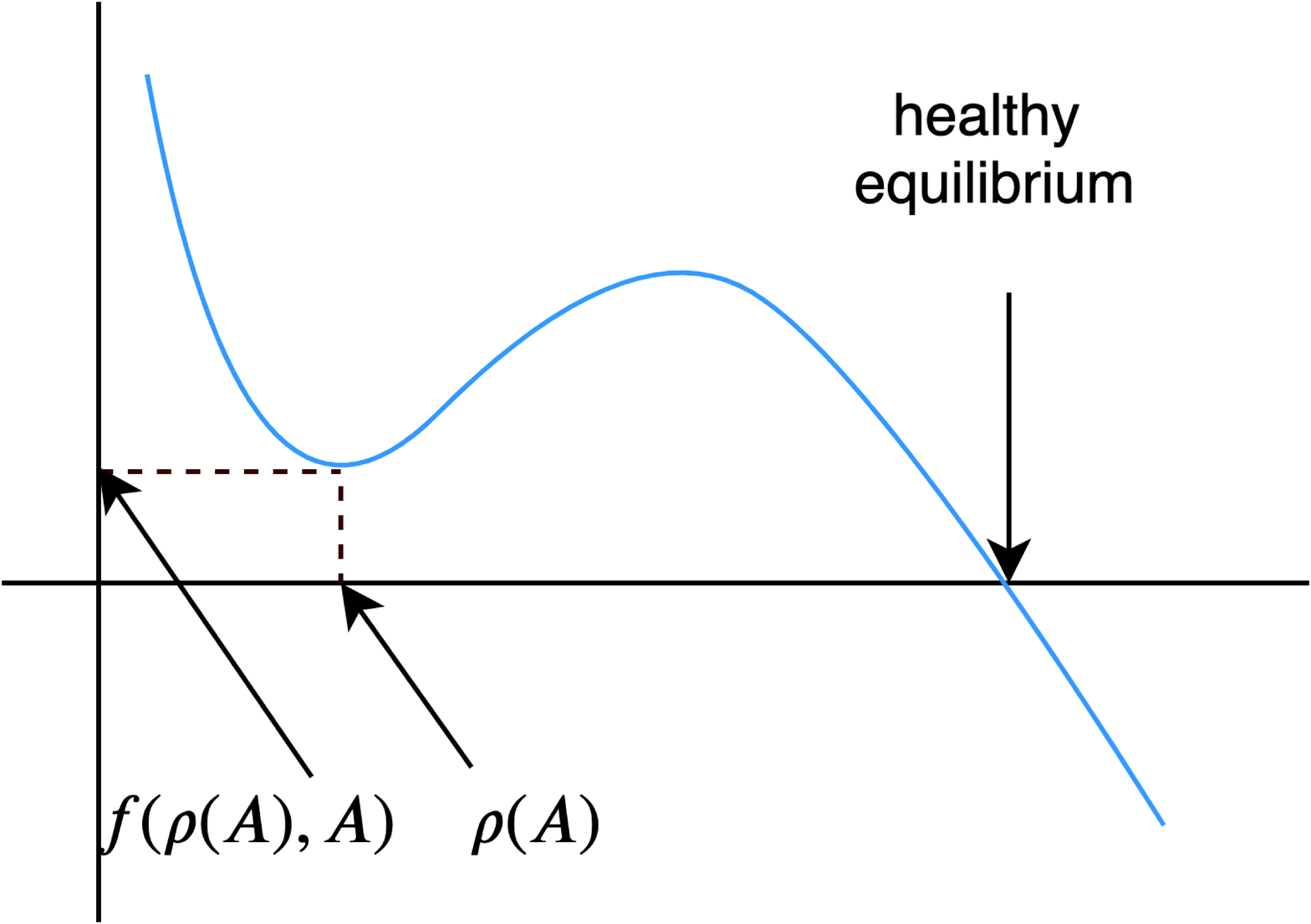}}}
\caption{In Figure 2(a) we can see a system before the saddle-node bifurcation, where both the unhealthy and the healthy equilibria are present. In Figure 2(b), we see a system after the saddle-node bifurcation, where the unhealthy equilibrium has been annihilated} \label{fig2}
\end{figure}

\quad We are in dynamics system $\dot{x}=f(x,a)$, with $f$ smooth. We will assume that this system always has a stable equilibrium $x_d>0$ that is not close to the origin and the saddle-node bifurcation can happen close to the origin, see Figure \ref{fig2}. Note that here $A$ denotes a vector of parameters and not just one. 

\quad The stable equilibrium away from the origin is a desirable state of the system and will it be called \textit{healthy}. The possible stable equilibrium close to the origin is an undesirable state of the system and it will be called \textit{unhealthy}. If in the system the unhealthy equilibrium is absent, then we say that the system is resilient.

\quad As is it can be seen from the Figure \ref{fig2} in order to detect whether the system is resilient or not, we can look at the value of the local minimum and check its sign. If it is negative, then we are in the case shown in Figure \ref{fig2}(a). If it is positive, then we are in the case shown in Figure \ref{fig2}(b). We do this by simply finding the smallest positive root of the equation $f^{'}(x, A)=0$, we will denote this by $f(\rho(A), A)$.

\subsection{Dynamics on graph}
\quad Real systems are usually composed of numerous components linked via a complex set of weighted, often directed, interactions(show in Figure \ref{fig:1}(b)). Let $G$ be a weighted directed graph of $n$ vertices and $m$ edges and let $M$ be its weighted adjacency matrix. Using $G$ we couple $n$ one-dimensional dynamical systems. The dynamics of each one-dimensional system is described by the differential equation $\dot{x} = f(x, \textbf{A})$, where $f$ is a smooth function and $\textbf{A}$ is a vector of parameters. The coupling term is described by a smooth function $g(x, y ,\textbf{B})$, where $\textbf{B}$ is a vector of parameters. The dynamics of the system is described by

\begin{equation}\label{equ5}
    \dot{x_i} = f(x_i, \textbf{A}_i)+\sum_{j=1}^n M_{ji} g(x_i,x_j, \textbf{B}_{ij})
\end{equation}

\quad We assume that the parameters of the equation (\ref{equ5}) are similar for every node but not exactly the same. We assume that each parameter is a random variable that gets a different realization on each node.

\quad We denote that $X={x_1,...,x_2}\in R^N$ and we define $F:R^N\to R^N$ by \begin{equation}\label{equ6}
    (F(\textbf{X},\textbf{A},\textbf{B}))_i=f(x_i,\textbf{A}_i)+\sum_{j=1}^n M_{ji}g(x_i,x_j,\textbf{B}_{ij})
\end{equation}
Then the system of equations (\ref{equ6}) can be written as\begin{equation}\label{equ7}
    \dot{\textbf{X}} = F(\textbf{X}, \textbf{A}, \textbf{B})
\end{equation}
The equilibrium of the system satisfies $F(\textbf{X}_e,\textbf{A},\textbf{B})=0$.

\quad Generally, we do not know very well when $\dot{\textbf{X}} = F(\textbf{X}, \textbf{A}, \textbf{B})$ will be resilient in a large-scale network. It is more difficult to know the resilience of $\dot{\textbf{X}}$ when considering uncertainty on parameters of vectors $\textbf{A}, \textbf{B}$ and uncertainty on topology (e.g. properties of $M_{ij}$) in dynamic network.

\section{Approach and Methodology}

\subsection{Dynamic network with uncertainty}
\quad Uncertainty in dynamic network may exit in self-dynamics of each component in $f(x,\textbf{A})$ and each component in coupling term $g(x,y, \textbf{B})$ as well as the network topology. We assume that each parameter is a random variable that gets a different realization on each node and moreover the value of any parameters has to be within a range of its true value. So we have $\textbf{A}=\textbf{A}(1+e_1U)$, $\textbf{B}=\textbf{B}(1+e_2U)$, $M=M(1+e_3U)$, where $U$ a random variable uniform in $[a, b]$ and $e_1, e_2, e_3$ constants. The mathematics model of dynamic network with uncertainty is showed as:
\begin{equation}\label{equ8}\begin{split}
    &\dot{x_i}=f(x_i,\textbf{A}_i(1+e_1U))+\\ &\sum_j^nM_{ji}(1+e_3U)g(x_i,x_j,\textbf{B}_{ij}(1+e_2U))
    \end{split}
\end{equation}

\subsection{Two-step method to estimate resilience with uncertainty}
\quad The first step is to use mean field dynamics and central limit theorem to get the expression which describes the probability of resilience of dynamic network. The second step is to use Polynomial Chaos Expression (PCE) to calculate the probability.

\subsubsection{Mean field dynamics}

\quad In order to find the mean field approximation of the equilibrium of the system, we define $\textbf{1}:={1,...1}\in R^N$
\begin{equation}\label{equ9}\begin{split}
    &\Xi :=Mean[F(x\textbf{1, A, B})] =\\ &\frac{1}{n}\sum_{i=1}^n(f(x,A_i))+\frac{1}{n}\sum_{i,j=1}^n M_{ji}g(x,x,B_i)
    \end{split}
\end{equation}

\quad Note that $\Xi(x)$ depends on \textbf{A} and \textbf{B}. Since \textbf{A} and \textbf{B} are random variables, for any $x$, $\Xi(x)$ also a random variable. Then we search for $r$ such that $\Xi(x)=0$.

\quad Because, the parameters $\textbf{A}_i$ are assumed to be iid random variables, for fixed $x$, $f(x, \textbf{A}_i)$ are also iid random variables. We define
\begin{equation}\label{equ10}
    \mu_{f(x)}:=\textbf{E}[f(x,\textbf{A}_i)]
\end{equation}
\begin{equation}\label{equ11}
    \delta_{f(x)}:=\sqrt{\textbf{Var}[f(x,\textbf{A}_i)]}
\end{equation}
This means that by Central Limit Theorem, for big enough $n$, $\frac{1}{n}\sum_{i=1}^nf(x,\textbf{A}_i)$ can be approximated by a normally distributed random variable with mean $\mu_{f(x)}$ and standard deviation $\frac{1}{n}\delta_{f(x)}$, i.e
\begin{equation}
    \frac{1}{n}\sum_{i=1}^nf(x,\textbf{A}_i)\sim\textbf{N}(\mu_{f(x)}, \frac{1}{n}\delta_{f(x)}^2)
\end{equation}\label{equ12}
Similarly, the random variables $g(x,x,\textbf{B}_{ij})$ are i.i.d, we define
\begin{equation}\label{equ13}
    \mu_{g(x)} := \textbf{E}[g(x,x,\textbf{B}_{ij})]
\end{equation}
\begin{equation}\label{equ14}
    \delta_{g(x)}:=\sqrt{\textbf{Var}[g(x,x,\textbf{B}_{ij})]}
\end{equation}
Then we have 
\begin{equation}\label{equ15}
    \frac{1}{n}\sum_{i,j=1}^n\textbf{M}_{ji}g(x,x,\textbf{B}_{ij})\sim\textbf{N}(\frac{m}{n}\mu_{g(x)},\frac{m}{n^2}\delta_{f(x)}^2)
\end{equation}

\quad For dynamic network with uncertainty, we define the auxiliary functions:
\begin{equation}\label{equ16}
    \phi(x, \textbf{U}) = f(x,\textbf{E}[\textbf{A}](1+e_1\textbf{U}))
\end{equation}

\begin{equation}\label{equ17}
    \varphi(x,\textbf{U},U)=\sum_j^n\textbf{E}[M](1+e_3U)g(x,x,\textbf{E}[\textbf{B}](1+e_2\textbf{U}))
\end{equation}
Let $k$ be the dimension of $\textbf{A}$ and $l$ be the dimension of $\textbf{B}$, then for the function $f$ we define
\begin{equation}\label{equ18}
    \mu_{f(x)}=\int\limits_{[a,b]^k}\frac{1}{(b-a)^k}\phi(x,\textbf{U})d\textbf{U}
\end{equation}
and 

\begin{equation}\label{equ19}
    \delta_{f(x)}^2=\int\limits_{[a,b]^k}\frac{1}{(b-a)^k}(\phi(x,\textbf{U})^2-\mu_{f(x)}^2)d\textbf{U}
\end{equation}
Similarly, for $g$ we define

\begin{equation}\label{equ20}
    \mu_{g(x)}=\int\limits_{[a,b]^{l+1}}\frac{1}{(b-a)^{l+1}}\varphi(x,\textbf{U},U)d\textbf{U}dU
\end{equation}
and

\begin{equation}\label{equ21}
     \delta_{g(x)}^2=\int\limits_{[a,b]^{l+1}}\frac{1}{(b-a)^{l+1}}(\varphi(x,\textbf{U},U)^2-\mu_{g(x)}^2)d\textbf{U}dU
\end{equation}

\quad Since $\Xi(x)$ is the sum of 2 normally distributed random variables, when we combine the above we get 
\begin{equation}\label{equ22}
    \Xi(x)\sim\textbf{N}(\mu_{f(x)}+\frac{m}{n}\mu_{g(x)},\frac{1}{n}\delta_{f(x)}^2+\frac{m}{n^2}\delta_{g(x)}^2)
\end{equation}
We can get a realisation of $\Xi_\alpha(x)$ by drawing $\zeta_\alpha$ from $\textbf{N}(0,1)$ and setting 
\begin{equation}\label{equ23}
    \Xi_\alpha(x) = \mu_{f(x)}+\frac{m}{n}\mu_{g(x)}+\sqrt{\frac{1}{n}\delta_{f(x)^2}+\frac{m}{n^2}\delta_{g(x)}^2}\zeta_\alpha
\end{equation}
We assume that every realisation of $\Xi(x)$ has the shape described in Figure \ref{fig2}, i.e. it is close to a saddle-node bifurcation. We find that the smallest positive root $\rho$ of $\Xi^{'}(x)$. Finally we set $\tau=\Xi(\rho)$.

\quad Since $\Xi(x)$ is a random variable, both $\rho$ and $\tau$ are random variables. Moreover, $\tau$ is an indicator for the saddle-node bifurcation. For a given realization of $\zeta_\alpha$, if $\tau_\alpha>0$, then there is only one equilibrium and the dynamics is resilient and if $\tau_\alpha<0$, then there are three equilibria and the dynamics is non-resilient. Thus the probability of the system being resilient is $\textbf{P}(\tau>0)$. We can use a Polynomial chaos expansion (PCE) truncated to degree $n$ to approximate $\tau(\zeta)$, we will denote this PCE by $\widetilde{\tau}_n(\zeta)$. We define the function 
\begin{equation}\label{equ24}
    pos(x)= \left\{
             \begin{array}{lr}
             1  \quad \textup{if} \quad x>0 &  \\
             0  \quad \textup{otherwise} &  
             \end{array}
\right.
\end{equation}
Then, the probability that the system is resilient is given by the integral
\begin{equation}\label{equ25}
    \frac{1}{\sqrt{2\pi}} \iint\limits_{-\infty}^{+\infty}pos(\widetilde{\tau}_n(\zeta))\,d\zeta
\end{equation}

\subsubsection{Polynomial chaos expansion}
\quad Let $\Xi$ be random variable with known probability distribution function (PDF) $w$. Moreover let $X=\phi(\zeta)$, with $\phi$ a function that is square integrable on $\textbf{R}$ with $w$ as weight function, let us call this space $L_w^2$. Our goal is to approximate $X$ by a polynomial series of $\zeta$.

\quad For this we need a family of polynomials $P_n$ such that $P_0$ is not 0, for all $n$ the polynomial $P_n$ has degree $n$ and are orthogonal with respect to $w$, i.e. the inner product
\begin{equation}\label{equ26}
    <P_n, P_m>_w = \int_{-\infty}^{+\infty}P_m(x)P_n(x)w(x)\,dx
\end{equation}
is 0 when $m\ne n$. Moreover we assume that $P_0$ is normalized so that $<P_0,P_0>_w = 1$. The polynomials $P_n$ can be used as a basis for $L_w^2$. So we can write
\begin{equation}\label{equ27}
    \phi(\zeta) = \sum_{n\ge0}a_nP_n(\zeta)
\end{equation}
In order to get the expression of $\phi(\zeta)$, we need to define the orthogonal basis $P_{n}$ and the coefficients $a_{n}$. What kind of orthogonal basis should be chosen depends on the distribution of random variable $\zeta$. If random variable $\zeta$ obeys a Gaussian distribution, we can choose the Hermite polynomial as the orthogonal basis. If random variable $\zeta$ obeys uniform distribution, we can choose Legendre polynomial as the basis (shown in Table \ref{tab1})\cite{ref19}.

\begin{table}
\caption{Correspondence of the type of orthogonal basis to the type of random variable}
{\begin{tabular}{llll} 
\hline
&random variable & orthogonal basis & support\\
 \hline
 Continuous & Gaussian & Hermite & $(-\infty, \infty)$  \\
 & Gamma & Laguerre & $[0, \infty)$   \\
 & Beta & Jacobi & [a, b]  \\ 
 & Uniform & Legendre & [a, b]\\
 Discrete & Poisson & Charlier & {0, 1, 2...}\\
 & Binomial & Krawtchouk & {0 ,1, 2...}\\
 & Negative binomial & Meixner & {0, 1, 2...}\\
 & Hypergeometric & Hahn & {0, 1, 2...}\\
 \hline
\end{tabular}}
\label{tab1}
\end{table}

\quad Because $P_n$ is an orthogonal basis, we can get the coefficients by projecting on each basis vector
\begin{equation}\label{equ28}
    a_n = \frac{<\phi, P_n>_w}{<P_n,P_n>_w}
\end{equation}

\quad In order to do any computation with a PCE series, we need to truncate it. First, we notice that if the series converges, then the size of each coefficient goes to 0 if we take the limit of any index to infinity. This means that for every convergent such series we can ignore terms of order higher than some $N$. However for a given problem it is not trivial to find which exactly this $N$ is. Usually this is done by trial and error, where we can calculate more terms until the size of the new terms is smaller than the precision we need.

\quad For the computation of the coefficient we will use a non-intrusive method. We start by truncating the series to an arbitrary order $N$, $\phi_n(\zeta)=\sum_{n=0}^N a_nP_n(\zeta)$ and assume that this is enough for the wanted precision. Then we observe that this is a linear relation with respect to $a_n^{'}$. So we generate $M>N$ instances of the random variable $\zeta$,${\zeta_1,\zeta_2,...,\zeta_M}$. Then for every $\zeta_i$ we have the equation 
\begin{equation}\label{equ29}
    \phi(\zeta_i) = \sum_{n=0}^N a_n P_N(\zeta_i)
\end{equation}
Notice that $\phi(\zeta_i)$ and $P_n(\zeta_i)$ are just numbers and now we can compute the coefficients $a_n$ by solving a linear regression. After that we compute $\textup{sup}_\zeta|a_NP_N(\zeta)|$ and if it is smaller than the precision we stop, otherwise we increase $N$ and repeat the process.

\section{Results}
\subsection{Case study: mutualistic dynamics}
\quad We will apply the above method in the case of mutualistic dynamic on a graph. We set
\begin{equation}\label{equ30}
    f(x, B, C, K) = B+x(\frac{x}{C}-1)(1-\frac{x}{K})
\end{equation}
\begin{equation}\label{equ31}
    g(x,y,D,E,H) = \frac{xy}{D+Ex+Hy}
\end{equation}
where B, C, K, D, E and H are positive parameters.  We assume that some of them are random variables that get different realization on each node. We set $\textbf{E}(B)=0.1$, $\textbf{E}[C]=1$, $\textbf{E}[D]=5$, $\textbf{E}[K]=5$, $E=0.9$, $H=0.1$. We moreover assume that the value of any parameter has to be within $10\%$ its mean, so we have $B=\textbf{E}[B](1+0.1U)$, $C=\textbf{E}[C](1+0.1U)$ and so on, where $U$ a random variable uniform in $[-1, 1]$.

\begin{figure}
\centering
\subfloat[graph of function $\Xi_{\alpha}$ ]{%
\resizebox*{8cm}{!}{\includegraphics{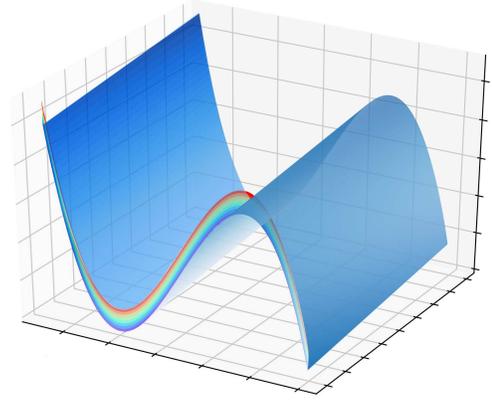}}}\hspace{5pt}
\subfloat[$\zeta_{\alpha}$ has different values]{%
\resizebox*{8cm}{!}{\includegraphics{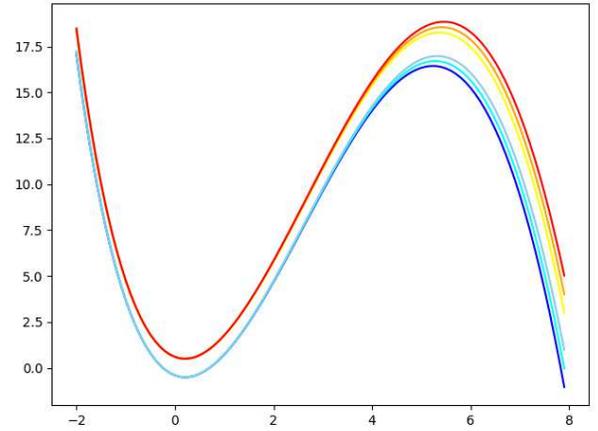}}}
    \caption{(a)graph of function $\Xi_{\alpha}(x)$(b)Graph of function $\Xi_{\alpha}(x)$ projects to XZ plane. When $\zeta_{\alpha}$ has different values, graphs of function $\Xi_{\alpha}(x)$ are different and the smallest positive root $\rho$ are different. Whether the system is resilient could be estimated through the figure.}
\label{fig3}
\end{figure}

\quad We define auxiliary functions
\begin{equation}\label{equ32}\begin{split}
    &\phi(x,U_1,U_2,U_3)=f(x,\textbf{E}[B](1+0.1U_1),\\
    &\textbf{E}[C](1+0.1U_2),
    \textbf{E}[K](1+0.1U_3))
    \end{split}
\end{equation}
and
\begin{equation}\label{equ33}
    \varphi(x,U_4,U_5)=\frac{\textbf{E}[M](1+0.1U_5)x^2}{\textbf{E}[D](1+0.1U_4)+Ex+Hx}
\end{equation}
Then for the function $f$ we define
\begin{equation}\label{equ34}
    \mu_{f(x)}:=\iiint\limits_{[-1,1]^3}\frac{1}{8}\phi(x, U_1, U_2,U_3)\,dU_1\,dU_2\,dU_3
\end{equation}
and
\begin{equation}\label{equ35}
    \delta_{f(x)}^2:=\iiint\limits_{[-1,1]^3}\frac{1}{8}(\phi(x, U_1,U_2, U_3)^2-\mu_{f(x)}^2)\,dU_1\,dU_2\,dU_3
\end{equation}
Similarly for $g$ we define
\begin{equation}\label{equ36}
    \mu_{g(x)}:=\iint\limits_{[-1,1]^2}\frac{1}{4}\varphi(x, U_4, U_5)\,dU_4\,dU_5
\end{equation}
and 
\begin{equation}\label{equ37}
    \delta_{g(x)}^2:=\iint\limits_{[-1,1]^2}\frac{1}{16}(\varphi(x,U_4,U_5)^2-\mu_{g(x)}^2)\,dU_4\,dU_5
\end{equation}
According to the above method, we can get a realisation of $\Xi_\alpha(x) = \mu_{f(x)}+\frac{m}{n}\mu_{g(x)}+\sqrt{\frac{1}{n}\delta_{f(x)}^2+\frac{m}{n^2}\delta_{g(x)}^2}\zeta_\alpha$. The figure of the function $\Xi_\alpha(x)$ is shown in Figure \ref{fig3} when $\zeta_\alpha$ has different values.

So we can see that every realisation of $\Xi(x)$ has the shape described in Figure \ref{fig2}. We can then find the smallest positive root $\rho$ of $\Xi^{'}(x)$, then use PCE to approximate $\tau({\zeta}$). 

\subsection{Convergence test of PCE}
\begin{figure}[htbp]
\centering
\resizebox*{8cm}{!}{\includegraphics{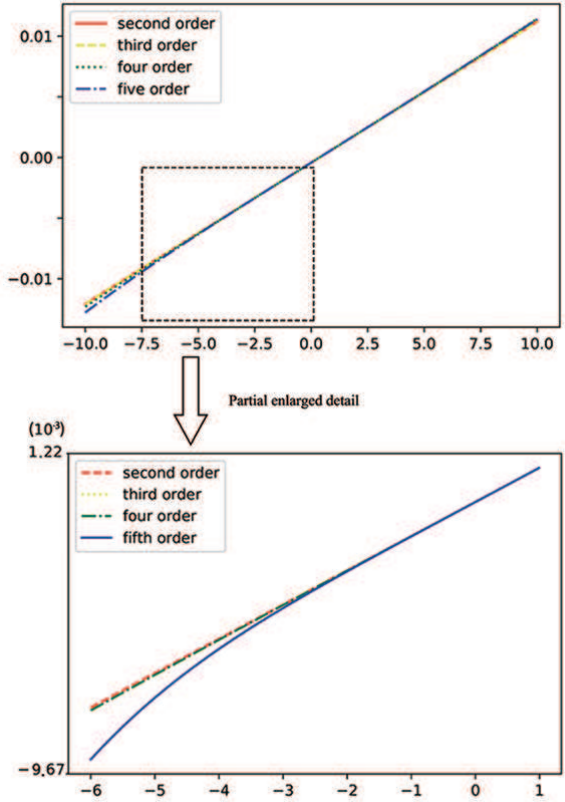}}
\caption{Approximate $\tau(\xi)$ by Hermite Polynomials. }
\label{fig4}
\end{figure}

\quad Since $\zeta$ obeys Gaussian distribution, we choose Hermite polynomial as orthogonal basis (shown in Table \ref{tab2}). We truncate the series to arbitrary orders $N$ from 2 to 5 shown in Figure \ref{fig4}. Increasing the order ($N$) of the polynomial improves the convergence of the function. However, increasing the order of the polynomial means that a substantially higher number of simulations is required. Therefore, a compromise between accuracy and required computational time is necessary. 

\begin{figure}[htbp]
\centering
\resizebox*{8cm}{!}{\includegraphics{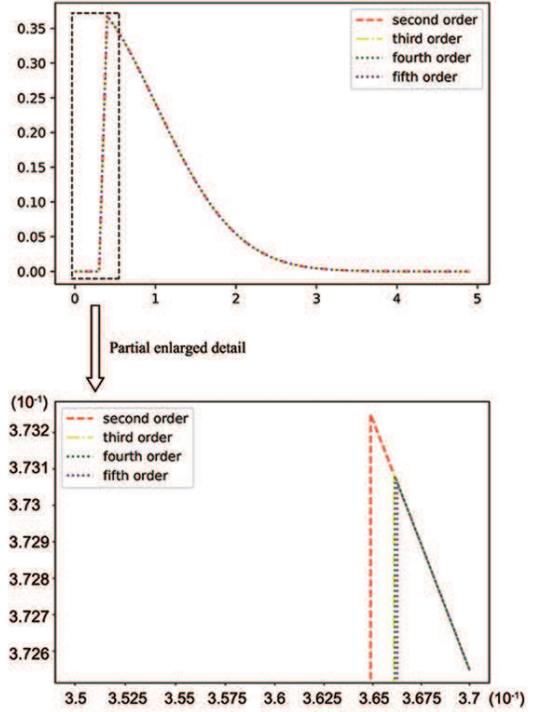}}
\caption{According to the PCE of $\tau(\xi)$, we can get the PDF of resilience of the system.}
\label{fig5}
\end{figure}

\begin{figure}[htbp]
\centering
\resizebox*{8cm}{!}{\includegraphics{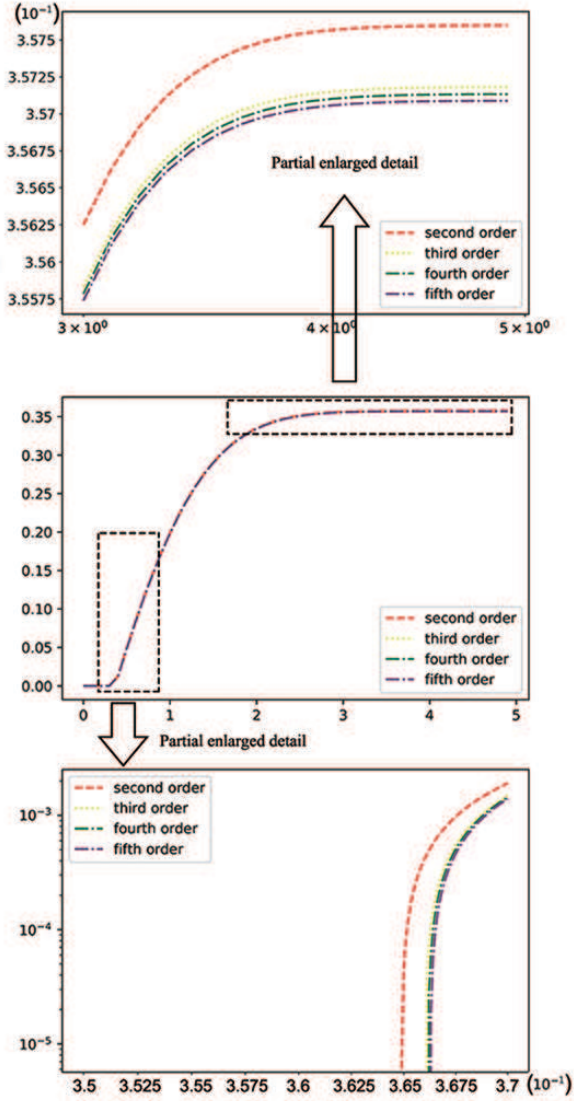}}
\caption{Get CDF of resilience of system by PDF}
\label{fig6}
\end{figure}

\quad Reference to the graph in Figure \ref{fig4}, it is impossible to infer which order of $N$ yields sufficient convergence of the PCE process. According to PCE in Figure \ref{fig4}, we can get the PDF with different truncation order in Figure \ref{fig5}. We can easily find the difference among different order especially $N=2$. In order to estimate the probability of resilience, we obtain a graph of Cumulative Distribution Function (CDF) with different truncation in Figure \ref{fig6}. It can be seen that the results for $N=3, N=4, N=5$ almost overlap while there is significant difference for $N=2$ in comparison to $N=3$.

\quad Therefore, $N=3$ can be considered as the appropriate choice for the polynomial order since choosing higher order polynomials substantially increases the required simulation time with only minor effects on improving the accuracy of the results.

\begin{table}
\centering
\caption{Hermite Polynomials}
{\begin{tabular}{lll} 
\hline
sequence number & Probability & Physics\\
 \hline
 $H_0(x)$ & 1 & 1  \\
 $H_1(x)$ & x & 2x   \\
 $H_2(x)$ & $x^2-1$ & $4x^2-2$  \\ 
 $H_3(x)$ & $x^3-3x$ & $8x^3-12x$\\
 $H_4(x)$ & $x^4-6x^2+3$ & $16x^4-48x^2+12$\\
 $H_5(x)$ & $x^5-10x^3+15x$ & $32x^5-160x^3+120x$\\
\hline
\end{tabular}}
\label{tab2}
\end{table}

\begin{figure}[htbp]
\centering
\subfloat[]{
\resizebox*{8cm}{!}{\includegraphics{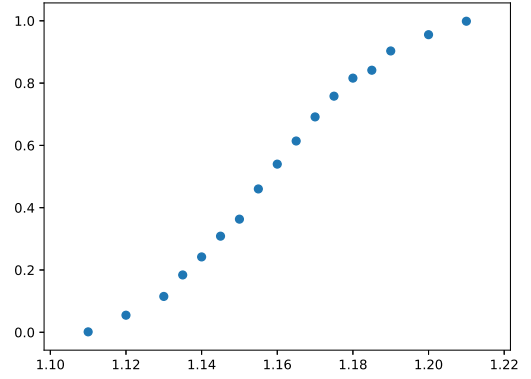}}}\hspace{5pt}
\subfloat[]{
\resizebox*{8cm}{!}{\includegraphics{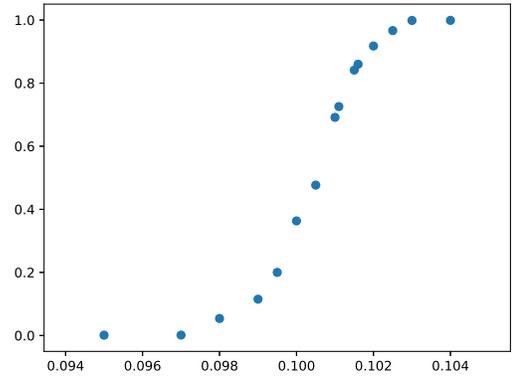}}}
\caption{(a) shows that probability of resilience is positive correlated to the weight of network. (b) shows that probability of resilience is positive correlated to the parameter $B$.} 
\label{fig7}
\end{figure}

\subsection{Analysis}
\quad In order to know how topology of network influence resilience of the system, we need to do parameter sensitivity analysis of the system, such as weight of edges. In Figure \ref{fig7}(a), we can see that probability of resilience is correlated to the weight of system. Strong connectivity promote resilience since the effect of perturbation are eliminated through inputs from the broader system. In mutualistic system, the first term on the right hand side of equation (\ref{equ30}) account for the incoming migration at a rate $B$ from neighbour ecosystems. The positive relationship between parameter $B$ and probability of resilience of mutualistic dynamic system (show in Figure \ref{fig7}(b) means that incoming migration from neighbour ecosystem could make this system more possible to be resilient.

\section{Conclusion and Future Work}
\quad Currently, we do not understand how to estimate resilience of dynamic networked systems with multiple model parameter uncertainty. In this paper, we built a mean-field informed Polynomial Chaos Expansion (PCE) model to quantify the uncertainty for a wide range of uncertainty distributions. This approach can effectively estimate the resilience behaviour of an arbitrarily large networked system and analyze the effect of both topological and dynamical parameters on the system. In the future, we will develop multi-resolution algorithms to achieve local to global resilience prediction.


%

.






%
{}

%










\end{document}